\documentclass{article}
\usepackage[utf8]{inputenc}
\usepackage{color}

\usepackage{latexsym}
\usepackage{amsmath}
\usepackage{amsfonts}
\usepackage{amssymb}
\usepackage{amsthm}
\usepackage{amscd}
\usepackage{mathrsfs}

\begin{document}
\newcommand{\nn}{\nonumber\\}
\renewcommand{\theequation}{\arabic{section}.\arabic{equation}}

                                        %
\def\si{\sigma}                         %
\def\lm{\lambda}                        %
\def\bC{{\mathbb C}} 
 
\def\be{\begin{equation}}              %
\def\ee{\end{equation}}                %
                                        %

\input ArtNouvc.fd
\newcommand*\initfamily{\usefont{U}{ArtNouvc}{xl}{n}}

\def\no{\noindent}                      %
\def\om{\omega}
\def\nl{\nabla^L}
\def\ups{\Upsilon}
\def\ba{\bar a}
\def\bchi{\bar\chi}
\def\omt{\tilde{\omega}}
\def\ti{\tilde}
\def\bR{{\mathbb R}}                    %
\def\bC{{\mathbb C}}  
\def\tum{\Tr{(VM)}}
\def\bN{{\mathbb N}}    
\def\bcj{{\mathbb C}^{1\vert 1}}
\def\bZ{{\mathbb Z}}                    %
\def\bT{{\mathbb T}} 
\def\bma{\begin{pmatrix}}
\def\ema{\end{pmatrix}}
\def\A{{\cal A}}
\def\F{{\cal F}}
\def\o{\Omega}
\def\bchi{\bar\chi^i}
\def\In{{\rm Int}}
\def\ba{\bar a}
\def\w{\wedge}
\def\ep{\epsilon}
\def\k{\kappa}
\def\slmn{sl(m\vert n,\bC)}
\def\sumn{su(m\vert n)}
\def\sld{sl(2\vert 1,\bC)}
\def\sud{su(2\vert 1)}
\def\osprs {osp(2r\vert s,\bC)}
\def\osp{osp(2\vert 1,\bC)}
\def\uosp{uosp(2\vert 1)}
\def\Uosp{UOSp(2\vert 1)}
\def\st{\stackrel{\otimes}{,}}
\def\Tr{{\rm tr}}
\def\TTr{{\rm Tr}}
\def\ST{{\rm str}}
\def\STr{{\rm STr}}
\def\ss{\subset}
\def\ga{\gamma}
\def\bga{\bar\gamma}
\def\bc{{\bf C}}
\def\br{{\bf R}}
\def\de{\delta}
\def\sus{S^{2\vert 2}} 
\def\tr{\triangleleft}
\def\al{\alpha}
\def\la{\langle}
\def\stvm{\ST(\V\M)}
\def\ra{\rangle}
\def\G{{\cal G}}
\def\dd{\ddagger}
\def\th{\theta}
\def\bth{\bar\theta}
\def\lm{\lambda}
\def\jp{\frac{1}{2}}
\def\js{\frac{1}{4}}
\def\d{\partial}
\def\dz{\partial_z}
\def\dbz{\partial_{\bar z}}
\def\db{\partial_b}
\def\dbb{\partial_{\bar{b}}}
\def\1{{\mbox{\boldmath $1$}}}  

\def\be{\begin{equation}}
\def\ee{\end{equation}}
\def\bea{\begin{eqnarray}}
\def\eea{\end{eqnarray}}
\def\apo{C^{pol}(S^2)}
\def\sapo{C^{pol}(S^{2\vert 2})}
\def\B{{\cal B}}
\def\M{{\cal M}}
\def\U{{\cal U}}
\def\cP{{\cal P}}
\def\C{{\cal C}}
\def\R{{\cal R}}
\def\V{{\cal V}}
\def\S{{\cal S}}
\def\g{{\cal G}}
\def\H{{\cal H}}
\def\E{{\cal E}}
\def\Q{{\cal Q}}
\def\agh{(\A,\g,\H)}
\def\tro{d\mu_{S^2}}
\def\std{d\mu_{S^{2\vert 2}}}
\def\bT{\bar{\cal T}}
\def\Z{{\cal Z}}
\def\si{\sigma}
\def\*{\ddagger}
\def\j{\dagger}
\def\Ad{{\rm Ad}}
\def\vv{\vec{\times}}
\def\ri{{{\mathrm{i}}}}    
\def\vph{\vec{\Phi}}
\def\vps{\vec{\Psi}}
\def\bz{\bar{z}}
\def\e{\varepsilon}
\def\b{\beta}
\def\tt{{\bf\times}}
\def\ot{\otimes}
\def\bb{\bar b}
\def\ri{\rm i}
\def\De{ (\epsilon V)}
\def\bk{\bar k}
\def\deg{\de^{\g}}
\def\sg{*_{\g}}
\def\vko{V_{\H^{\perp}}}      %
\def\bph{{\bf \Phi}}
\def\bps{{\bf \Psi}}
\def\bu{{\bf u}}

\def\cW{{\cal W}}                       %
\def\fR{{\mathfrak{R}}}
\def\fS{{\mathfrak{S}}}
\def\bep{\bar\epsilon}

\def\stw{\stackrel{\wedge}{,}}
\def\cM{{\cal M}}                       %
\def\cC{{\cal C}}                       %

\def\cK{{\cal K}}                       %

\newtheorem{theorem}{Theorem}[section]
\begin{titlepage}
\vspace*{0.5cm}
\begin{center}
{\Large \bf  }

\end{center}

\vspace{0.2cm}

\begin{center}
{\large \bf  On supermatrix models, Poisson geometry and  noncommutative supersymmetric gauge theories}\\ 
[50pt]{\small
{ \bf Ctirad Klim\v{c}\'{\i}k}
\\
Aix Marseille Universit\'e, CNRS, Centrale Marseille\\ I2M, UMR 7373\\ 13453 Marseille, France}
\end{center}

\vspace{0.2cm}

\begin{abstract} We  construct a  new supermatrix model which represents a manifestly supersymmetric noncommutative regularisation of the $UOSp(2\vert 1)$ supersymmetric Schwinger model on the supersphere.  Our construction is much
simpler than those already existing in the literature and it was found by using Poisson geometry in a substantial way. 
 
  \end{abstract}

\end{titlepage}
\newpage

\section{Introduction}
The subject of this paper is situated on a crossroad of two active  current themes of research: the first concerns the application
of the method of localisation  \cite{Pestun} to extract  quantitative informations about rigidly supersymmetric
Euclidean  field theories  on compact manifolds  and the second deals with the study of (super)matrix models giving rise to noncommutative field theories with various amount of (super)symmetry.
It is certainly impossible to provide a complete bibliography of relevant works in both directions we can however mention  papers \cite{BC,BPZ,CC,DGFL,DG,FS,GGK,GF,GL,HL,HR,SaS,Sh,ST} resp. \cite{CW,IKTW,IU,IU2, Kl2,Ste,Yd} which treat the gauge theories living on two-dimensional compact Euclidean (super)spaces, mostly on a sphere $S^2$  or on a  supersphere $\sus$.

Probably the first example of a  supersymmetric gauge theory on the supersphere  was defined and studied in  \cite{Kl1},    namely, the $UOSp(2,1)$ invariant supersymmetric extension of the standard Schwinger model on $S^2$ \cite{Sch,Jay}. This rigidly supersymmetric model yields in the decompactification limit the minimal $N=(1,1)$ supersymmetric electrodynamics on the plane constructed already by Ferrara in \cite{Fer} and it was studied in \cite{Kl1}   with
the main motivation to find its manifestly supersymmetric noncommutative regularisation keeping only a finite number of degrees of freedom in the theory. This goal was indeed achieved in \cite{Kl1} and the resulting theory even flaunted a solid geometrical status in both commutative and noncommutative version. In particular, it allowed also a formulation in terms of a supermatrix model
as discovered later  in \cite{IU,IU2}. Inspite of all of those successes here we   come back to the subject to present another construction which accomplishes precisely the same objectives as\cite{Kl1} but is at the same time  much simpler.
The basic ingredient making possible to simplify the story of \cite{Kl1} is the use of Poisson geometry which not only allows to guess natural candidates for $\uosp$ supersymmetric gauge invariant Lagrangians but it 
incredibly streamlines and speeds up the technical work needed to verify that they have the required properties. 
We invented and tested this new Poisson formalism  while studying supersymmetric $\sigma$-models \cite{Kl3} and we
are pleased to confirm its efficiency in the present work.

Let us thus straightaway write down two principal results of the present article. The first one is
a new very simple version of the action principle of the $UOSp(2\vert 1)$ supersymmetric Schwinger model on the supersphere $S^{2\vert 2}$. It reads
\be  S( \Phi,\E)= \int\std \left( \vert\vert \{\cM^2,\Phi\}+[\cM,\E]\Phi\vert\vert^2+\frac{1}{4e^2}\vert\vert \{\cM^2,\ST\{\cM^2,\E\}\}\vert\vert^2 \right).\label{fc}\ee
The second one is a new action of the supermatrix model which in the limit of the infinite size of the supermatrices yields the
theory \eqref{fc}. It is given by
\be S_N(\hat\Phi,\cP)=\alpha_N\STr ~\ST\biggl(( \cP\hat\Phi-\hat\Phi\hat\cK)^\dagger(\cP\hat\Phi-\hat\Phi\hat\cK)
-\frac{\sigma_N^2}{e^2}[ \cP, \ST(\cP^2_v\cP^2_v-\hat\cK^2_v\hat\cK^2_v)]^2\biggr),\label{fb}\ee
where the  $N$-dependence of the constants $\alpha_N$ and $\sigma_N$ is made explicit in \eqref{fsa} and \eqref{two}.

The Introduction is not a place where all technical details should be given, nonetheless we believe that it is helpful to provide the reader with a rough acquaintance with the notations met
in \eqref{fc} and \eqref{fb} already at this level. Thus $\Phi$ stands for the charged matter superfield on $\sus$ and the $(N+1\vert N)\times (N+1\vert N)$ supermatrix $\hat\Phi$ is its noncommutative analogue.
$\E$ is the so called superspinorial  $(2\vert 1)\times (2\vert 1)$ supermatrix  the entries of which are functions on $\sus$ and
it plays the role of the gauge superfield on the supersphere.  $\cP$ is the noncommutative analogue of $\E$, it
is the superspinorial   $(2\vert 1)\times (2\vert 1)$ supermatrix  the entries of which are $(N+1\vert N)\times (N+1\vert N)$ supermatrices. The bracket $\{.,.\}$ is the Kirillov-Costant-Souriau Poisson bracket\footnote{ This Kirillov-Costant-Souriau Poisson bracket is constructed in the standard way by viewing   the supersphere   as the coadjoint orbit of the supergroup $UOSp(2\vert 1)$.}  on $\sus$ and the  supervectorial $(2\vert 1)\times (2\vert 1)$ supermatrix  $\cM$  is the moment map generating the infinitesimal $\uosp$ transformations of $\sus$ via these Poisson brackets. Finally, $\hat\cK$ is the noncommutative analogue of $\cM^2$. 

It must be stressed that the compact  notation appearing in \eqref{fc} and \eqref{fb}  was not conceived forcibly
but it is very natural. We mean by this that one can perform  nontrivial technical operations on our actions working directly in the succint notation  without e.g. choosing
a basis of the Lie superalgebra $\uosp$, without detailing the entries of the moment map $\cM$ or of the superspinorial
supermatrix $\E$ and, of course, without expanding the superfields in components. In particular, the   $\uosp$ superinvariance as well as the gauge invariance of the action \eqref{fc} can be checked in this concise way just by making use of some basic properties of the Poisson brackets like the Jacobi identity.  For that matter, we believe  
that the use of the Poisson geometry in the construction of the supersymmetric invariants will prove to be useful also
for other compact supermanifolds enjoying rigid supersymmetry whenever the action of the corresponding Lie superalgebra
on the supermanifold is Hamiltonian.

The plan of the paper is as follows: in the bosonic warm up Section 2, we first  present purely bosonic counterparts \eqref{bos} and \eqref{bfe}  of the main results \eqref{fc} and \eqref{fb}. In Section 3.1, we review the basic properties of the supermatrices and of the supersphere, in Section 3.2 we construct the action
\eqref{fc} and establish its symmetry properties, in Section 3.3, we show that 
  the expansion of the supersymmetric action \eqref{fc} in components contains the purely bosonic action
\eqref{bos} as well as the standard Schwinger model on the sphere \cite{Sch, Jay}. We review the concept of the
fuzzy supersphere in Section 3.4 and in Section 3.5 we finally construct the supermatrix model \eqref{fb}, we establish its symmetry properties and we prove that in the large $N$ limit it yields the  supersymmetric gauge theory \eqref{fc}. The last Section 4 is devoted to a discussion of the results. It should also be read carefully since
we formulate there some interesting geometrical questions that
our concept of the supergauge field poses.

\setcounter{equation}{0}

\section{Bosonic warm up}
\subsection{Manifestly $SO(3)$ symmetric  electrodynamics on $S^2$} The standard way for writing down 
the action of a scalar  electrodynamics living on a two-dimensional Riemannian space-time $M$ with Euclidean signature uses the
determinant $\sqrt{g}$ of the Riemann tensor $g_{\mu\nu}$ and the components $g^{\mu\nu}$ of it inverse:
\be S(\phi,A_\mu)=\jp\int_M d^2\xi \sqrt{g} \biggl( g^{\mu\nu}(\d_\mu\bar\phi+iA_\mu\bar\phi)(\d_\mu-iA_\mu)\phi +\frac{1}{e^2}g^{\mu\rho}g^{\nu\sigma}F_{\mu\nu}F_{\rho\sigma}\biggr).\label{ga}
\ee
Here $A_\mu$ is the electromagnetic potential in some coordinates $\xi^\mu$, $e^2$ the coupling constant  and $F_{\mu\nu}$ is the field strength:
\be F_{\mu\nu}=\d_\mu A_\nu-\d_\nu A_\mu.\ee
Of course, a potential term $\int\tro V(\bar\phi\phi)$ can be obviously added 
to this action but we shall be systematically avoiding it as our principal concern is  the interaction of the matter field $\phi$ with the gauge field $A_\mu$.

If the manifold $M$ is the unit two-sphere $S^2$ equipped with the standard round Riemannian metric then the action \eqref{ga} 
can be rewritten\footnote{Of course, some work is needed to recover \eqref{ga} from \eqref{act} which we leave to the reader. To do that, the most straightforward way is to view $S^2$ as the Riemann sphere and to use the standard complex coordinate $z$ to parametrize the complement of the north pole.} in a $SO(3)$-covariant way as
\be S(\phi, A_k) = \jp\int\tro \biggl((R_k+iA_k)\bar\phi(R_k-iA_k)\phi+ \frac{1}{e^2}F_k(A)F_k(A)\biggr),\label{act}\ee
where the $SO(3)$ covariant electromagnetic field strength vector $F_k$ is defined as
\be F_k(A):=\epsilon_{klm}(R_lA_m-R_mA_l+\epsilon_{lmp}A_p).\label{sof}\ee
In order to explain the meaning of the symbols $A_k$ and $R_k$, $k=1,2,3$ we  first need  to introduce three
functions $x_1,x_2,x_3$ on $S^2$ the values of which in every point of the sphere are given by the $\bR^3$ Cartesian coordinates
of this point (the unit sphere $S^2$ is thought to be standardly embedded in the three-dimensional Euclidean space $\bR^3$ 
and the Riemannian metric on $S^2$ is induced from the flat one on $\bR^3$). Thus it holds 
\be x_1^2+x_2^2+x_3^2=1\label{sur}\ee
 and the measure on $S^2$ induced by the round Riemannian metric can be written accordingly as
\be \tro=dx_1dx_2dx_3\delta(x_1^2+x_2^2+x_3^2-1).\ee
The following vector fields $R_k$ on $\bR^3$ 
are tangent to the surface \eqref{sur} of the embedded sphere hence they can be viewed also as the vector fields on $S^2$:
\be R_k:=-\epsilon_{klm}x_l\partial_m.\label{vec}\ee
The vector fields $R_k$ generate an infinitesimal action of $SO(3)$ on $S^2$ and are related by an obvious identity following from the total antisymmetry of the Levi-Civita tensor $\epsilon_{klm}$:
\be x_kR_k=0.\label{ors}\ee
To define $A_k$, we decompose the electromagnetic potential $1$-form $A_\mu d\xi^\mu$ as
\be A_\mu d\xi^\mu= A_k\beta_k, \label{dec}\ee
where 
\be \beta_k:=-\epsilon_{klm}x_ldx_m.\ee
If we furthermore impose a constraint  
\be A_kx_k=0\label{tc}\ee
then the coefficient functions $A_k$   are determined uniquely from $A_\mu d\xi^\mu$ via \eqref{dec} and \eqref{tc}. 
The constraint \eqref{tc} has a natural geometric interpretation because it says that the $SO(3)$ vector $A_k$
is perpendicular to the normal vector $x_k$, hence $A_k$ is tangent to the surface of the sphere.

Virtually all authors working on the subject of the  gauge theories on the fuzzy sphere
\cite{BV,CW,IKTW,Ki,Kl2,KNP,Ste,StS} used the  manifestly $SO(3)$ invariant form \eqref{act2} of the scalar electrodynamics on $S^2$  as the starting point
to the construction of noncommutative deformations.  It appears that nothing more can be said about \eqref{act}, yet there is an almost "banal" detail which we remarked only recently and which actually triggered our  renewed interest in the subject of the fuzzy deformations of (super)gauge field theories. The point is that the action \eqref{act}
can be rewritten in slightly different  but still manifestly $SO(3)$-invariant way as follows
\be S(\phi, A_k) = \jp\int\tro \biggl((R_k+iA_k)\bar\phi(R_k-iA_k)\phi+ \frac{1}{e^2}F(A)^2\biggr),\label{act2}\ee
where 
\be F(A):=\epsilon_{klm}x_kR_lA_m.\label{sfs}\ee
The quantity $F(A)$ can be referred to as a "scalar field strength" and it is invariant with respect to the
$SO(3)$ transformations (infinitesimaly generated by the vector fields $R_k$) as well as with respect to the gauge transformations 
\be A_k\to A_k+R_k\rho.\label{gtr}\ee
Here $\rho$ is arbitrary function on $S^2$. Notice  also, that the gauge transformation \eqref{gtr} is compatible with the constraint \eqref{tc} due to \eqref{ors}.

\medskip

\noindent {\bf Remark 1}: {\small
In order to demonstrate the equivalence of the actions \eqref{act} and \eqref{act2}, it is useful
to parametrize the electromagnetic potential $ A_\mu d\xi^\mu$ in terms of its Hodge dual $*(A_\mu d\xi^\mu)$:
\be *(A_\mu d\xi^\mu):=B_k\beta_k, \label{dec2}\ee
where, as before, the coefficient functions $B_k$ are unambiguously fixed by the constraint 
\be B_kx_k=0.\label{tc2}\ee
It is easy to check that the relation between the respective $SO(3)$ covariant parametrizations $B_k$ of $*(A_\mu d\xi^\mu)$ and $A_k$ of $A_\mu d\xi^\mu$
reads
\be B_k=-\epsilon_{klm}x_lA_m, \qquad A_k=\epsilon_{klm}x_lB_m.\label{Hod}\ee
It is now easy to calculate the quantities $F_k$ and $F$ in  the dual $B_k$ parametrization. The result is
\be F_k=x_k(R_mB_m), \qquad F =R_mB_m,\ee
from which the equivalence of the actions \eqref{act} and \eqref{act2} readily follows.}

\medskip

Is there any conceptual or technical gain which could be extracted from the rewriting \eqref{act2} of the action \eqref{act}? Well, as we shall see in Section 2.2, the noncommutative deformation based on the new version
\eqref{act2}  looks more or less as complicated as the standard one based on  \eqref{act}.   
 However, the things are very different in the supersymmetric setting (cf. Section 3.2 further on) where the  existence of a scalar super field strength  generalizing the purely bosonic quantity $F$  turns out to simplify drastically the construction of the noncommutative supersymmetric gauge theory.  The reason for this is the following:  in the $\Uosp$ supersymmetric analogue of the action \eqref{act} constructed
in \cite{Kl1} the role of the three-component  field strength $F_k$ is played by an eigth-component object living
in the adjoint representation of the supergroup $SU(2\vert 1)$. This eight-component object has to be, furthermore, constrained in
the $\Uosp$ supersymmetric and supergauge invariant way. All in all, the already existing construction \cite{Kl1} of the manifestly supersymmetric electrodynamics
on the supersphere  is quite involved while, as we shall see in Section
3, it can be replaced by an astonishingly simple alternative, by using the scalar superfield strength as a "sesame". 

\subsection{The use of Poisson geometry}

 Before turning to the supersymmetric case, which is our real concern in this paper, 
we   spend here some more  time with  the purely bosonic $SO(3)$ invariant gauge theory \eqref{act2}. We do it  in order to illustrate 
the  use of the new technical
tools based on Poisson geometry.  As we have seen in Section 2.1, in the purely  bosonic case  the use of the Kronecker tensor and of the Levi-Civita tensor is sufficient to obtain all important formulas therefore the Poisson tools represent just an amusing computational alternative. However, in the supersymmetric case the Poisson tools are considerably more beneficial from both  conceptual as well as   technical point of view because  $su(2\vert 1)$ invariant tensors are more numerous than in $so(3)$ case, they have more components and  they are  tied together by more complicated identities.

The Poisson geometry enters the game because there is a natural Kirillov-Kostant-Souriau Poisson bracket\footnote{  The sphere $S^2$  can be viewed  as the coadjoint orbit of the group $SO(3)$.}   on $S^2$
 defined by
 \be \{x_i,x_j\}=\epsilon_{ijk}x_k,\label{PBS}\ee
 which allows to express the action of the rotation vector fields $R_k$ on the complex scalar field $\phi$  in a Hamiltonian way:
 \be R_k\phi =\{x_k,\phi\}.\label{epr}\ee
The bracket \eqref{PBS}  can be also described more invariantly if we introduce the so called moment map $M$. The latter is a
   traceless idempotent Hermitian $2\times 2$ matrix the entries of which are the complex functions $x_3, x_\pm:=x_1\pm{\ri} x_2$ on the sphere:
 \be M=\bma x_3 & x_-\\ x_+ & -x_3\ema. \label{mmp}\ee
 The set of the defining Poisson brackets \eqref{PBS} can be then rewritten in several equivalent ways
 which are useful in different contexts. For example
 \be \{\Tr(V_1M),\Tr(V_2M)\}=-{\ri}\Tr{([V_1,V_2]M)}\label{UV}\ee
 where $V_1,V_2$ are arbitrary 
 traceless Hermitian matrices representing the Lie algebra $so(3)\equiv su(2)$ in the spin $\jp$ representation (the choice of the Pauli matrices for $V_1,V_2$ gives readily \eqref{PBS}). Other descriptions of this fundamental Poisson structure on the sphere are matrix-like, e.g.
 \be \{\Tr(VM),M\}={\ri}[V,M],\label{soeq}\ee
 or
 \be\{M,M\}=2{\ri}M; \label{mmm}\ee
the last representation should be read in components as
 \be\sum_j \{M_{ij},M_{jk}\}=2{\ri}M_{ik}.\ee
 Finally it holds true also
 \be \{M_{ij},M_{kl}\}={\ri}\delta_{jk}M_{il}-{\ri}\delta_{il}M_{kj}.\label{pdr}\ee
 
 In what follows, it will be convenient to represent the electromagnetic potential $A_k$ also by a traceless Hermitian matrix
 $  A$:
\be    A\equiv\bma A_3&A_1-{\ri}A_2\\A_1+{\ri}A_2&-A_3\ema.\ee
By using  \eqref{epr}, the scalar field strength \eqref{sfs} can be then  written as 
\be F(A)=\frac{\ri}{2} M_{kl}\{M_{lp}, A_{pk}\} \equiv -\frac{\ri}{2} \Tr(M\{M, A\})\label{sfsb}\ee
and the new form of the manifestly $SO(3)$ invariant action \eqref{act2} can be recast as
\be S(\phi, A):= \js\int \tro \biggl( \vert\vert\{M,\phi\}-{\ri}  A\phi)\vert\vert^2 - \frac{1}{2e^2} \Tr^2(M\{M, A\})\biggr).\label{bos}\ee
Here
\be \vert\vert\{M,\phi\}-{\ri} A\Phi)\vert\vert^2:=\Tr\left ((\{M,\bar\phi\}+{\ri}  A\bar\phi)(\{M,\phi \}-{\ri} A\phi )\right).\ee
Consider the following infinitesimal variations of the fields
$\phi$ and $A$: 
\be \delta_{V}\phi:=\{\Tr(VM),\phi\}, \qquad  \delta_V A:=  -{\ri}[V, A] +\{\Tr(VM), A\}. \label{vara}\ee
Here an arbitrary element $V$ of $so(3)\equiv$Lie$(SO(3))$  is viewed as the traceless Hermitian matrix in the spin $\jp$ representation of  $so(3)$; in particular the choice of the Pauli matrices $V=\sigma_k/2$ gives
\be \delta_k\phi:=\{x_k,\phi\}=R_k\phi, \qquad  \delta_kA_l:=  \epsilon_{kml}A_m +\{x_k,A_l\}. \label{vara2}\ee
Notice also that the  $so(3)$ variation \eqref{vara2} of $A$ is induced from the Lie derivative of the potential $1$-form  $A_\mu d\xi^\mu$: \be \delta_k(A_\mu d\xi^\mu)\equiv (\iota_{R_k}d+d\iota_{R_k})(A_\mu d\xi^\mu).\ee
 We now wish to show that the action \eqref{bos} is $so(3)$ invariant  with respect to
 to  the variations \eqref{vara} of the interacting fields $\phi$ and $   A$:
 \be \delta_VS\equiv S(\phi+\delta_V\phi, A+\delta_V  A)-S(\phi, A)= 0.\label{nvar}\ee
 To do that we use Eq. \eqref{soeq} and we calculate
 \be \{\Tr(VM), \{M,\phi\}\}= {\ri }[V,\{M,\phi\}]+\{M,\{\Tr(VM),\phi\}\}\label{ppp}\ee
 and
 $$ \{\tum , \Tr{(M\{M,A\})}\} =$$ $$=\Tr\biggl({\ri }[V,M]\{M,A\}+ M\{{\ri}[V,M],A\}+M\{M,\{\tum,A\}\}\biggr)=$$
 \be =  \Tr\biggl(-M\{M,{\ri }[V,A]\}+M\{M,\{\tum,A\}\} \biggr).\label{qqq}\ee
From \eqref{ppp},\eqref{qqq}  and \eqref{vara} we then derive
\be \{M,\delta_V\phi\}\}=\{\Tr(VM), \{M,\phi\}\}-{\ri }[V,\{M,\phi\}];\label{rdm}\ee
\be \Tr{(M\{M,\delta_V A\})}=\{\tum , \Tr{(M\{M,A\})}\}.\label{rdmb}\ee
The relations \eqref{rdm} and \eqref{rdmb}  imply
\be \delta_VS =\js\int\tro\{\tum,  \vert\vert\{M,\phi\}-{\ri}  A\phi)\vert\vert^2 -  \jp\Tr^2(M\{M, A\})\}.\label{pem}\ee
We finish the proof of the fact that $\delta_VS=0$  by   exploiting the $so(3)$ invariance of the measure $\tro$:
 \be 0=\delta_V (\int\tro f) =
 \int\tro\delta_Vf=\int\tro\{\Tr(VM),f\}.\label{bid}\ee
The relation \eqref{bid} holds for any function $f$   on $S^2$, in particular for that appearing in \eqref{pem}.
 
Now we verify the gauge invariance of the action $S(\Phi,A)$ with respect to the following gauge transformation depending on an arbitrary function $\rho$ on $S^2$:
\be  A\to A+\{M,\rho\}, \quad \phi\to e^{\ri \rho}\phi.\label{gau}\ee
The check of the invariance of the matter kinetic term is trivial, however a little bit more work is needed to establish the invariance of the
field strength:

Because the moment map $M$ squares to the unit matrix, we  obtain, respectively, for every function $f$ on $S^2$ and
for every matrix function $T$ on $S^2$
\be 0=\Tr\{M^2,f\}=\Tr(M\{M,f\}+\{M,f\}M)=2\Tr (M\{M,f\});\label{nula}\ee
\be 0=\Tr\{M^2,T\}=\Tr(M\{M,T\}) -\Tr(M\{T,M\}).\label{jeden}\ee
The relations \eqref{nula}, \eqref{mmm}  and also \eqref{jeden}, considered for $T=\{M,f\}$, then imply
\be 2\Tr(M\{M,\{M,f\}\})= \Tr(M\{\{M,M\},f\}=2{\ri}\Tr(M\{M,f\})=0.\label{tri}\ee
Obviously, the equation \eqref{tri} guarantees the invariance of the field strength \eqref{sfsb} 
with respect to the gauge transformation \eqref{gau}. For that matter,
we should perhaps recall that the field $A$ is constrained by
the constraint $x_kA_k=0$ which can be rewritten also as
\be \Tr(M A)=0.\label{con1}\ee
We observe from \eqref{nula}, that  this constraint is also preserved by the gauge transformation \eqref{gau} as it should.

\medskip

\no {\bf Remark 2}: {\small There is an  alternative way of writing  the action
\eqref{bos} in terms of the Hodge dual fields $B_k$ given by \eqref{Hod}. Introducing the traceless Hermitian matrix $B$ by
\be    B\equiv\bma B_3&B_1-{\ri}B_2\\B_1+{\ri}B_2&-B_3\ema,\ee
the relations \eqref{Hod} can be rewritten as
\be B:=\frac{\ri}{2}[M,A], \quad   A=-\frac{\ri}{2}[M, B]\label{Hod2}\ee
and the action \eqref{bos} as
\be S(\phi, B)= \js \int \tro \biggl(\vert\vert\{M,\phi\}-\jp{} [M, B]\phi\vert\vert^2 +  \frac{1}{2e^2}\Tr^2\{M, B\}\biggr).\label{bos2}\ee
In this dual way of writing the action of the scalar electrodynamics on the sphere
the kinetic term of the gauge field gets simpler at the price of rendering the matter kinetic term slightly more complicated.  
This alternative action \eqref{bos2} is of course also $so(3)$ invariant and 
gauge invariant  with respect to the following $so(3)$ and gauge transformations
of the  fields $\phi$ and $\hat B$:
\be \delta_{V}\Phi:=\{\Tr(VM),\Phi\}, \qquad   \delta_V B :=  -{\ri}[V, B] +\{\Tr(VM),B\};\ee
\be \Phi\to e^{\ri \rho}\Phi, \quad  B\to  B-\frac{\ri}{2}[M,\{M,\rho\}] \quad .\ee
Notice that these transformations respect the constraint $\Tr{(M B)}=0$.}

\subsection{Scalar electrodynamics on the fuzzy sphere}
Now we are going to present a new construction of a purely bosonic scalar electrodynamics on the fuzzy sphere $S^2_N$\cite{Ho1,Mad}. Recall  that $S^2_N$   is the noncommutative
manifold resulting from  the quantization of $S^2$ induced by the Poisson brackets \eqref{PBS}. The   linear $SO(3)$-equivariant quantization map $Q_N$   associates  to  smooth functions $f$ on $S^2$   sequences  
of $N\times N$-matrices  $Q_N(f)$ which are called the quantized or fuzzy functions.    We shall not need an explicit  formula for the  quantization map $Q_N$ but we do need three basic properties of it:
\be Q_N(f)Q_N(g)=Q_N(fg)+O\biggl(\frac{2}{\sqrt{N^2-1}}\biggr),\label{p1}\ee
\be [Q_N(f),Q_N(g)]= {\ri}\frac{2}{\sqrt{N^2-1}}Q_N(\{f,g\})+ O\biggl(\frac{4}{N^2-1}\biggr),\label{p2}\ee
 \be  \frac{1}{\pi}\int_{S^2}\tro f = \frac{2}{\sqrt{N^2-1}}\TTr (Q_N(f))+O\biggl(\frac{4}{N^2-1}\biggr).\label{p3}\ee
 Obviously the parameter $2/\sqrt{N^2-1}$ plays the role of the Planck constant for the quantization map $Q_N$.  It is also important to stress that   $\TTr$ stands for the trace of the $N\times N$ matrices while, throughout this paper, we reserve the symbol $\Tr$ 
for the trace of $2\times 2$ matrices.

 To give a flavor, what the map $Q_N$ is about, let us  make explicit  the quantized versions of the   functions  $1$, $x_3$,
$x_1\pm {\ri} x_2$  on $S^2$:
\be Q_N(1)=\1_N,\ee
\be Q_N(x_3)_{ij}=\frac{1}{\sqrt{N^2-1}}(N+1-2j)\delta_{ij},\ee
\be Q_N(x_1+ {\ri} x_2)_{ij}=\frac{2}{\sqrt{N^2-1}} \sqrt{(j-1)(N-j+1)}\delta_{i,j- 1} \label{fux}\ee 
\be Q_N(x_1- {\ri} x_2)=Q_N(x_1+ {\ri} x_2)^\dagger,\ee 
where $\1_N$ stands for the unit $N\times N$-matrix. In particular, it is then easy to verify that it holds the emblematic fuzzy sphere relation 
\be Q_N(x_1)^2+ Q_N(x_2)^2+ Q_N(x_3)^2=Q_N(1),\label{fuz}\ee

In what follows, we shall adopt a notation keeping the dependence on $N$ tacit:
\be \hat x_k:=Q_N(x_k).\ee
It can be easily checked that  the following Hermitian matrices $L_k$
\be L_k:=\frac{\sqrt{N^2-1}}{2}\hat x_k,\label{gn}\ee
realize an $N$-dimensional unitary representation of the Lie algebra $so(3)$. This fact is compatible with the property \eqref{p2} and with the definition \eqref{PBS}.

\medskip

Let us now construct a field theory on the fuzzy sphere, the fields of which are a complex $N\times N$ 
matrix $\hat\phi\equiv Q_N(\phi)$ and three Hermitian $N\times N$ matrices $\hat A_k\equiv Q_N(A_k)$ obeying a constraint
\be \hat x_k\hat A_k+\hat A_k\hat x_k +\frac{2}{\sqrt{N^2-1}} \hat A_k\hat A_k=0.\label{cnn}\ee
For the
action functional we take:
\be S_N(\hat \phi,\hat A_k)=\frac{\pi}{\sqrt{N^2-1}}\TTr\biggl(([L_k,\hat\phi]+\hat A_k\hat\phi)^\dagger([L_k,\hat\phi]+\hat A_k\hat \phi) -\frac{1}{e^2}F(\hat A)^2\biggr),\label{bfe}\ee
where the $N\times N$ matrix $F(\hat A)$ defined by
\be F(\hat A)=-{\ri}\frac{2}{\sqrt{N^2-1}}\epsilon_{klm}\biggl((L_k+\hat A_k)(L_l+\hat A_l)(L_m+\hat A_m)-L_kL_lL_m\biggr)\label{ffs}\ee
plays the role of the fuzzy field strength. 
The fuzzy action \eqref{bfe} is invariant with respect to the following $so(3)$ field variations
\be \delta_p\hat \phi:=-{\ri}[L_p,\hat \phi], \quad \delta_p \hat A_k= \epsilon_{plk}\hat A_l-{\ri}[L_p,\hat A_k]\ee
because it can be easily checked
that 
$$ \delta_p([L_k,\hat\phi]+\hat A_k\hat \phi)\equiv [L_k, \delta_p \hat\phi]+(\delta_p\hat A_k)\hat\phi+\hat A_k\delta_p\hat\phi=$$
\be =\epsilon_{plk}([L_l,\hat\phi]+\hat A_l\hat\phi]) -{\ri} [L_p,[L_k,\hat\phi]+\hat A_k\hat \phi]          \ee
and
\be\delta_pF(\hat A)=-{\ri}[L_p, F(\hat A)].\ee
The $so(3)$ invariance of the constraint \eqref{cnn} can be verified in a similar way.

Our fuzzy action \eqref{bfe}  as well as the constraint \eqref{cnn} can be easily checked to be also gauge invariant with respect to the following gauge transformation:
\be \hat\phi\to U\hat\phi, \quad \hat A_k\to U\hat A_kU^{-1}-[L_k,U]U^{-1}, \label{tra}\ee
where $U$ is an arbitrary unitary $N\times N$ matrix. Notice in particular, that the noncommutative field strength $F(\hat A)$ transforms under \eqref{tra} as
\be F(\hat A)\to UF(\hat A)U^{-1}.\ee
There exists a more compact way of writing the fuzzy action \eqref{bfe} in the spirit of the commutative action
\eqref{bos} but also in the spirit of the matrix model filosophy of Ref. \cite{Ste}. So, as in the commutative case,
we define $2\times 2$ matrices
\be   \hat A\equiv\bma \hat A_3&\hat A_1-{\ri}\hat A_2\\\hat A_1+{\ri}\hat A_2&-\hat A_3\ema, \qquad 
 \hat M\equiv\bma \hat x_3&\hat x_1-{\ri}\hat x_2\\\hat x_1+{\ri}\hat x_2&-\hat x_3\ema 
\ee
and
\be \hat E:=\hat M+\frac{2}{\sqrt{N^2-1}}\hat A.\ee
In terms of the matrix $\hat E$,  
 the action \eqref{bfe} can be expressed compactly as
\be S_N(\hat\phi,\hat E)=\frac{\pi}{2}\frac{\sqrt{N^2-1}}{2}\TTr\biggl(\Tr((\hat E\hat\phi -\hat\phi\hat M)^\dagger(\hat E\hat\phi -\hat\phi\hat M)) -\frac{N^2-1}{16e^2}\Tr^2(\hat E^3-\hat M^3)\biggr),\label{bfem}\ee
where $\Tr$ means the trace of $2\times 2$ matrices and $\TTr$ the trace of $N\times N$ matrices.

Our last task in this section is to show that in the limit of large $N$ the matrix model \eqref{bfe} and  \eqref{bfem} gives the scalar electrodynamics
\eqref{bos} on the ordinary sphere.  The large $N$ limit of the kinetic term is easy to establish  since the relations \eqref{p1},\eqref{p2} and \eqref{gn} directly give
\be [L_k,\hat\phi]+\hat A_k\hat\phi = Q_N({\ri}\{x_k,\phi\}) +A_k\phi) + O\biggl(\frac{2}{\sqrt{N^2-1}}\biggr).\ee
A little more work  is necessary to find the large $N$ limit of the field strength
\be F(\hat A)=Q_N(\epsilon_{klm}x_k\{x_l,A_m\}) +  O\biggl(\frac{2}{\sqrt{N^2-1}}\biggr).\label{rlm}\ee
The derivation of \eqref{rlm} is based on the same relations \eqref{p1},\eqref{p2} and \eqref{gn} as before  but also on the constraint \eqref{cnn} which 
itself can be written as
\be 0=\hat x_k\hat A_k+\hat A_k\hat x_k +\frac{2}{\sqrt{N^2-1}} \hat A_k\hat A_k=2Q_N(x_kA_k)+O\biggl(\frac{2}{\sqrt{N^2-1}}\biggr).\ee
Finally, in the limit of large $N$  the trace $\TTr$ approaches the integral over the sphere following the relation \eqref{p3} and this is the last
ingredient needed to establish the correct $N\to\infty$ limit of the
fuzzy action \eqref{bfe}.

\setcounter{equation}{0}

\section{Fuzzy supersymmetric Schwinger model}
\subsection{Supermatrices and the supersphere $\sus$}

By an even $(2\vert 1)\times (2\vert 1)$ supermatrice over a complex Grassmann algebra we mean a square matrix $\V$
of the block form
\be \V=\bma A&B\\C&D\ema, \ee 
where $A$ is a $2\times 2$ square matrix with even entries,       the column vector $B$ and the row vector $C$ have odd entries and
$D$ is an even element of the Grassmann algebra.
In what follows, we shall exclusively consider the Grassmann algebras equipped with the so called  graded involution. The latter was introduced in \cite{RS}
and satisfies \be \overline{ab}=\bar a\bar b, \quad \overline{\bar a}=(-1)^{{\rm deg}(a)}a.\ee
By definition, the Hermitian conjugated matrix $\V^\dagger$ has entries which fulfil
\be (\V^\dagger)_{ij}:= {\rm sign}(i-j)\bar \V_{ji},\label{sun}\ee
 where  ${\rm sign}(i-j)$ takes value $1$ if $i\geq j$ and $-1$ otherwise. Every Hermitian $(2\vert 1)\times (2\vert 1)$ supermatrix $\V^\dagger =\V$  can be   unambiguously represented as a  sum  \be \V= \ST(\V)\1+\V_v+\V_s, \label{dcp}\ee where the supertrace   is defined by
\be \ST(\V):= \Tr(A)-D\ee
and $\V_v$,$\V_s$ are  traceless Hermitian supermatrices of   the respective forms
 \be \V_v= \bma v_3&\bar v&-\bar\nu\\ v &-v_3 & \nu\\ \nu &\bar\nu & 0\ema,\quad \V_s=\bma v_0& 0 &\vartheta\\ 0 &v_0 & -\bar\vartheta\\ \bar\vartheta&\vartheta & 2v_0\ema.\label{pro}\ee
 Now it can be easily checked that the ordinary commutator of two even traceless Hermitian supermatrices of the $v$-type\footnote{A traceless Hermitian supermatrix $A$  is of the $v$-type if $A_v=A$ or, equivalently, if $A_s=0$.} is again an even traceless Hermitian supermatrix of the $v$-type. The even traceless Hermitian
 supermatrices   of the $v$-type thus form a Lie superalgebra referred to as $\uosp$. It can be also verified that the commutator
 of the matrix of the  $v$-type with the matrix of the  $s$-type is of the $s$-type, therefore the space of the matrices of the $s$-type is the representation of $\uosp$ called the superspinorial representation.
 
The supersphere $\sus$ is a supermanifold generated by three even
real variables $y_k$, $k=1,2,3$ and a pair of graded complex conjugated odd variables $\th,\bth$  fulfilling one constraint
\be y_1^2+y_2^2+y_3^2+2\th\bth=1.\ee
It will prove extremely useful to organize the generators $y_k, \th,\bth$ into the even traceless Hermitian supermatrix of the $v$-type as follows:
   \be \cM=\bma y_3 &y_1-{\ri}y_2 & -\bth\\ y_1+{\ri}y_2& -y_3 & \th\\
    \th & \bth& 0\ema. \label{ort}\ee
 The $\uosp$ Kirillov-Kostant-Souriau Poisson structure on $\sus$ is then  defined by the  following bracket.  
\be \{\ST(\V_1\M),\ST(\V_2\M)\}:=-{\ri} \ST([\V_1,\V_2]\M), \quad \V_1,\V_2\in\uosp.\label{spb1}\ee
Thus the Hamiltonian generating the action of $\V$ on the supersphere $\sus$ via the Poisson bracket is $\ST(\V\M)$  and hence  $\M\in \uosp$ is nothing but
  the moment map of this action.  Other description of this fundamental Poisson structure on the supersphere are matrix-like, e.g.
 \be \{\stvm,\cM\}={\ri}[\V,\cM],\label{ssoeq}\ee
 or
 \be\{\cM,\cM\}=\frac{3}{2}{\ri}\cM. \label{smmm}\ee
  Although we do not need it, we list for completeness the Poisson brackets of the generators $y_k,\th,\bth$ as they follow  from the general formula \eqref{spb1}:
  $$\{y_k,y_l\}=\epsilon_{klm}y_m,\quad \{y_3,\th\}=-\frac{\ri}{2}\th, \quad \{y_3,\bth\}=+\frac{\ri}{2}\bth,$$
  $$\{y_1+{\ri}y_2,\bth\}=-{\ri} \th, \quad \{y_1-{\ri}y_2,\th\}=-{\ri} \bth, \quad \{y_1+{\ri}y_2,\th\}=0=\{y_1-{\ri}y_2,\bth\},$$
\be  \{\th,\th\}=-\frac{\ri}{2}(y_1+{\ri}y_2), \quad \{\bth,\bth\}=+\frac{\ri}{2}(y_1-{\ri}y_2), \quad \{\th,\bth\}= \frac{\ri}{2}y_3.\label{pba}\ee
 Unlike the $so(3)$ moment map $M$ considered in the previous section, the $\uosp$ moment map $\cM$ given by \eqref{ort} is 
 not idempotent. In fact, the square of $\cM$ is nontrivial and plays a very important role in the construction of the supersymmetric invariants.   Thus we define
 \be \cK:=\cM^2-2\label{mk1}\ee
 and we find that $\cK$ is  the Hermitian supermatrix  of the $s$-type. For completeness, we describe it explicitly:
 \be \cK=\bma w&0&\zeta\\0&w&-\bar\zeta\\\bar\zeta&\zeta&2w\ema,\label{mk2}\ee
 where
 \be \zeta:=\th (y_1-{\ri}y_2) -\bth y_3, \quad
 \bar\zeta :=\th y_3+\bth (y_1+{\ri}y_2), \quad w=-1-\th\bth.\label{rrr}\ee
 We finish this section by listing useful formulae
 expressing the Poisson brackets involving the matrix
 $\cK$:
 \be \{\cK,\cK\}=-  \frac{{\ri }}{2}\cM, \quad \{\cK,\cM\}=\{\cM,\cK\}=\frac{{\ri }}{2}\cK.\label{ufo}\ee

 \subsection{Supersymmetric electrodynamics on $\sus$}

By using the supermatrix $\cK$  defined by \eqref{mk1},\eqref{mk2}, a manifestly $\uosp$ invariant action of a free massless complex scalar superfield $\Phi$ on the supersphere was written in \cite{Kl3}. It reads
 \be S_0(\Phi)=\ST\int \std\{\cK,\bar\Phi\}\{\cK,\Phi\},\label{fr}\ee
 where the $\uosp$ invariant measure on the supersphere $S^{2\vert 2}$ is defined by \cite{GKP,Kl3} 
\be \std := dy_1dy_2dy_3 d\th d\bth \delta(y_1^2 +y_2^2+ y_3^2 +2\th\bth-1).\ee
 The immediate consequence of the invariance of the measure $\std$ is the formula
\be \int \std \{\M,f\}=0,\label{ui}\ee
valid for all functions $f$ on the supersphere (cf. \eqref{bid}). Thanks
to \eqref{ui} and to the following  identity
\be \{\cK,\{\stvm,\Phi\}\}=\{\stvm,\{\cK,\Phi\}\}-{\ri}[\V,\{\cK,\Phi\}]\label{aaa}\ee
derived from \eqref{ssoeq} and from \eqref{mk1}, the free action \eqref{fr} can be easily checked to be invariant with respect to the $\uosp$ action on $\Phi$ defined by
\be \delta_{\V}\Phi:=\{\stvm,\Phi\}.\ee
Let us gauge the global $U(1)$ symmetry $\Phi\to\exp{{\ri }\varrho_0}\Phi$ of the action \eqref{fr}
by introducing a multiplet  of odd and even gauge superfields $C,\bar C,C_0$ arranged in the matrix of the $s$-type:
\be \C:=\bma C_0&0&C\\0&C_0&-\bar C\\
\bar C&C&2C_0\ema.\ee
We require, moreover, that $\C$ is constrained by
\be \ST(\cK\C)=0.\label{conx}\ee
Consider now the following action in which the $s$-type matrix superfield $\C$ is viewed as external
 \be S_{ext}(\Phi,\C)=\ST\int \std(\{\cK,\bar\Phi\}+{\ri}\C\bar\Phi)(\{\cK,\Phi\}-{\ri}\C\Phi).\label{fre}\ee
 It is easy to check the symmetry of the action $S_{ext}$ with respect to the following gauge transformations
 \be \Phi\to\e^{{\ri}\varrho}\Phi, \quad \C\to \C+\{\cK,\varrho\},\label{sgt}\ee
 where $\varrho$ is an arbitrary even real function on $\sus$.

Now we are going to concentrate to the problem
 how to render the supermatrix gauge superfield $\C$ dynamical. Said in other words, we must construct a viable manifestly supersymmetric  kinetic for the gauge superfield $C$  to be added to the supersymmetric action \eqref{fre}.   One way of solving this problem was shown in \cite{Kl1}, the other, and drastically simpler,  is presented in this paper.  Why  we have not seen
 the simpler solution while writing the former paper \cite{Kl1}? Well, we were not aware at that time of a possibility to use Poisson geometry as a very efficient conceptual and technical tool for constructing superinvariant Lagrangians.   We have first used this tool only recently in the context of supersymmetric $\sigma$-model \cite{Kl3} and the present article
 constitutes another proof of its efficiency. Thus we introduce here a concept of a scalar superfield strength $\F(\C)$ 
 defined as  
 \be \F(\C):=\jp\ST(\cM\{\cK,\C\} +\{\C,\cK\}\cM).\label{ssfs}\ee
{\small  For completeness, we detail here this formula in terms of  the
 constituent superfields $C_0,C,\bar C$ of $\C$:
 $$\F(\C)= \bth(\{\bar\zeta,C_0\}-\{w,\bar C\})+\th(\{w,C\}-\{\zeta,C_0\})+$$\be +y_3(\{\zeta,\bar C\}+\{\bar\zeta, C\})+(y_1+{\ri}y_2)\{\zeta,C\}-(y_1-{\ri}y_2)\{\bar\zeta,\bar C\}.\ee}
  This scalar superfield strength $\F(\C)$ is really scalar which means that its variation  $\delta_{\V}\F$ induced by the vector-like transformation  \be \delta_{\V}\C:= -{\ri}[\V,\C]+\{\stvm,\C\}\ee 
  is simply 
 \be \delta_{\V}\F(C)= \{\stvm,\F(C)\}.\label{spf}\ee
 Let us prove the formula \eqref{spf} to give an illustration of the efficiency of our compact notation using the supermatrices $\cM$ and $\cK$. First of all, 
 since $\cK=\cM^2-2$, we infer from \eqref{ssoeq} that
 \be \{\stvm,\cK\}={\ri}[\V,\cK].\label{zzz}\ee
 Then we find from \eqref{ssoeq} and \eqref{zzz} 
$$\{\stvm,\ST(\cM\{\cK,\C\}\}=\ST({\ri}[\V,\cM]\{\cK,\C\})+\ST(\cM\{{\ri}[\V,\cK],\C\})+$$ $$+\ST(\cM\{\cK,\{\stvm,\C\}\}) =\ST({\ri}[\V,\cM\{\cK,\C\}]) -\ST(M\{\cK,{\ri}[\V,\C]\})+$$ \be +\ST(\cM,\{\cK,\{\stvm,\C\}\} =\ST(\cM\{\cK,\delta_\V\C\}).\label{11}\ee
Much in the same way, we find
\be \{\stvm,\ST(\{\C,\cK\}\cM)\}=\ST( \{ \delta_\V\C ,\cK\}\cM).\label{22}\ee
Eqs. \eqref{11} and \eqref{22} then clearly imply \eqref{spf}.

Let us now prove the invariance of the scalar superfield strength   $\F(\C)$ with respect to the gauge transformation \eqref{sgt}. By using the Jacobi identity in two alternative forms
\be \{a,\{b,\rho\}\}=\{\{a,b\},\rho\}+\{\{\rho,a\},b\};\ee
\be \{\{\rho,b\},a\}=\{\rho,\{b,a\}\}+\{b,\{a,\rho\}\},\ee
valid for arbitrary functions $a,b,\rho$ on $\sus$ with $\rho$ even,
we infer from \eqref{ufo}
\be \ST (\cM\{\cK,\{\cK,\rho\}\}-\{\{\rho,\cK\},\cK\}\cM )=
\ST (\cM\{\{\cK,\cK\},\rho\} )= -  \frac{{\ri }}{2}\ST(\cM\{\cM,\rho\}).\label{333}\ee
Because the supermatrix $\cM^2$ has constant supertrace (cf. \eqref{mk1},\eqref{mk2}), it holds
\be \ST(\cM,\{\cM,\rho\}=\jp\ST\{\cM^2,\rho\}=0.\label{444}\ee
Inserting \eqref{444} into \eqref{333}, we obtain the gauge invariance of $\F(\C)$, hence also the gauge invariance of the following action 
 \be S(\Phi,\C)=\ST\int \std\biggl((\{\cK,\bar\Phi\}+{\ri}\C\bar\Phi)(\{\cK,\Phi\}-{\ri}\C\Phi)-\frac{1}{e^2}\{\cK,\F(\C)\}^2 \biggr).\label{mca}\ee
The relations \eqref{aaa} and \eqref{spf} then easily imply also the $\uosp$ superinvariance of \eqref{mca}.

The compact and elegant expression \eqref{mca} for the action of the $\uosp$ supersymmetric Schwinger model on the supersphere constitutes the first main result of this article.  

\medskip

\no{\bf Remark 3}: {\small The supersymmetric Schwinger model on the supersphere can be rewritten in terms of  a dual parametrization of the gauge superfield $\C$ much in the
spirit of Remark 1. Thus define a new $s$-type traceless Hermitian matrix superfield by
\be \E:=-{\ri}[\cM,\C].\ee
Because of the constraint \eqref{conx}, it holds also
 \be \C:={\ri}[\cM,\E]\ee
 and
 \be \ST( \cK\E)=0.\ee
 The duality $\C\leftrightarrow\E$ may be called the Hodge one
 by analogy with  the purely bosonic case albeit we are
 not aware of its possible interpretation in the language of differential forms. A recent paper \cite{CCG} may
 possibly shed more light on this issue. Finally, the scalar superfield strength $\F$ in terms of $\E$ reads simply
 \be \F(\E)=-\frac{\ri}{2}\ST\{\cK,\E\}\label{dsf}\ee
 and the supersymmetric Schwinger model action becomes
\be S(\Phi,\E)= \ST\int \std\biggl((\{\cK,\bar\Phi\}-[\cM,\E]\bar\Phi)(\{\cK,\Phi\}+[\cM,\E]\Phi)-\frac{1}{4e^2}\{\cK,\ST\{\cK,\E\}\}^2 \biggr).\label{mcad}\ee
 It is in this form that we have presented the manifestly $\uosp$ supersymmetric action of the Schwinger model on $\sus$
 in the Introduction.}

\subsection{Component expansions}

In this section, we shall work out the action of the supersymmetric electrodynamics on the supersphere in components. We do it starting from the dual formulation \eqref{mcad} 
in which the scalar supersymmetric field strength has simpler form. Recall that the
gauge field $\E$ is  the Hermitian supermatrix of the type $s$:
 \be \E=\bma E_0&0&E\\0&E_0&-\bar E\\\bar E&E&2E_0\ema,\label{ctn}\ee
 which verifies the constraint
 \be \ST(\cK\E)=0.\label{nnn}\ee
 The constraint \eqref{nnn}  allows to express the even  superfield $E_0$
 in terms of the (mutually graded conjugated) superfields $E,\bar E$ as follows
 \be E_0=\bar\zeta E-\zeta \bar E.\label{eo} \ee
 The scalar superfield strength \eqref{dsf} in terms of the constituent superfields
$E_0$,$E$ and $\bar E$ becomes
\be \F(\E)={\ri}\{w,E_0\}-{\ri}\{\zeta,\bar E\}+{\ri}\{\bar\zeta,E\},\label{blf}\ee
or, by using the formula \eqref{eo}, as 
\be \F(\E)=\jp(\th E-\bth\bar E)+{\ri}(1+\th\bth)\left(\{\bar\zeta,E\}-\{\zeta,\bar E\}\right).\label{bfl2}\ee
Here $\th,\bth$ are related to $\zeta,\bar\zeta$ as in \eqref{rrr}. 

For the component expansions of the superfields $E,\bar E$ compatible with the graded conjugation we choose an ansatz
 \be E= {\ri} \zeta \tilde A_3 -{\ri}\bar\zeta (\tilde A_1-{\ri}\tilde A_2)+2(\bar \xi (y_1-{\ri}y_2)+\xi y_3)\zeta\bar\zeta,\label{bce}\ee
   \be \bar E= -{\ri} \bar\zeta \tilde A_3 -{\ri}\zeta (\tilde A_1+{\ri}\tilde A_2)+2(\bar\xi y_3-\xi (y_1+{\ri}y_2))\zeta\bar\zeta,\label{bcf}\ee
where  the real even components fields $\tilde A_k$ as well as they  mutually conjugated odd colleagues $\xi$ and $\bar\xi$, depend just on the variables $y_k$. Obviously, the component ansatz \eqref{bce},\eqref{bcf}
is not the most general one, since e.g. the zero order terms in $\zeta$,$\bar\zeta$ expansion are missing but all missing terms can be restored by a gauge transformation $\E\to\E -{\ri}[\M,\{\cK,\Lambda\}]$
for an appropriate choice of the even superfunction $\Lambda$. Thus the ansatz \eqref{bce},\eqref{bcf} is nothing but a variant of the Wess-Zumino gauge. 

The evaluation of the full scalar superfield strength $\F(\E)$ \eqref{bfl2} for the ansatz \eqref{bce},\eqref{bcf} gives:
\be \F(\E)={\ri}y_k\tilde A_k -\xi\bar\zeta -\bar\xi \zeta  +{\ri}\left(F(\tilde A)-y_k\tilde A_k+y_l\d_l(y_k\tilde A_k)\right)\zeta\bar\zeta,\label{dfs}\ee
where $F(\tilde A)$ is nothing but the purely bosonic scalar curvature \eqref{sfs}
\be F(\tilde A):=\epsilon_{klm}y_k\{y_l,\tilde A_m\}.\label{sfs2}\ee

Knowing \eqref{dfs}, we can now easily complete the evaluation of the component expansion of the gauge kinetic term in the action
\eqref{mcad}. It reads
 $$ \ST\int \std  \{\cK,\F(\E)\}^2  = -2\int \std \biggl(2\{\bar\zeta,\F(\E)\}\{\zeta,\F(\E)\}+\{w,\F(\E)\}^2\biggr)=$$$$= -4\int \std(1+\zeta\bar\zeta)  \{\bar\zeta,\F(\E)\}\{\zeta,\F(\E)\})=$$
 \be = - \int\tro \left((F(A)+ \rho)^2 +\{y_k,\rho\}\{y_k,\rho\}+ {\ri}\Xi^\dagger\sigma_k\{y_k,\Xi\}+\Xi^\dagger\Xi\right),\label{gki}\ee
  Here $\sigma_k$ are the Pauli matrices and  $\Xi$, $\Xi^\dagger$, $\rho$ et $A_k$ are defined as 
  \be \Xi:=  \bma \xi \\\bar\xi\ema, \qquad \Xi^\dagger =\bma \bar\xi &-\xi\ema,\ee
  \be \rho:=y_k\tilde A_k, \quad  A_k:=\tilde A_k-y_k\rho.\ee
  Note that $\rho$ and $A_k$ are, respectively, the radial and the tangential part of the field $\tilde A_k$ and $F(A)$ stands for the scalar curvature of the tangential part.

 We now proceed to the component expansion of the matter kinetic term. By using \eqref{ort},\eqref{mk2}, \eqref{ctn} and \eqref{nnn},  we  find successively
  $$ \ST\int \std\biggl((\{\cK,\bar\Phi\}-[\cM,\E]\bar\Phi)(\{\cK,\Phi\}+[\cM,\E]\Phi)\biggr)=$$
  $$= -2\int\std\biggl((\{\bar\zeta,\bar\Phi\}+{\ri }\bar C\bar\Phi)(\{\zeta,\Phi\}-{\ri }C\Phi)-(\{\zeta,\bar\Phi\}+{\ri } C\bar\Phi)(\{\bar\zeta,\Phi\}-{\ri } \bar C\Phi)+$$$$+(\{w,\bar\Phi\}
  +{\ri }C_0\bar\Phi)(\{w, \Phi\}-{\ri }C_0\Phi)\biggr)= -2\int\std(1+\th\bth)\times$$
  \be \times \biggl((\{\bar\zeta,\bar\Phi\}+{\ri }\bar C\bar\Phi)(\{\zeta,\Phi\}-{\ri }C\Phi)-(\{\zeta,\bar\Phi\}+{\ri } C\bar\Phi)(\{\bar\zeta,\Phi\}-{\ri } \bar C\Phi)\biggr).\label{vr}\ee
  Here $C={\ri}[\cM,\E]$, or, in detail:
  $$  C_0=-{\ri}\bth \bar E+{\ri}\th E$$
  \be \bar C=-{\ri}(1+\th\bth)(y_3\bar E+(y_1+{\ri}y_2)E), \quad C={\ri}(1+\th\bth)(y_3E-(y_1-{\ri}y_2)\bar E).
 \label{zh}\ee
  For the component expansion of the complex matter superfield $\Phi$, we choose the following ansatz (cf. \cite{GKP}):
  \be \Phi=\phi+\bth\psi_++\th\psi_-+(F+y_k\d_k\phi)\th\bth,\label{cme}\ee
\be \bar\Phi=\bar\phi+\bth\bar\psi_--\th\bar\psi_++(\bar F+y_k\d_k\bar\phi)\th\bth.\label{cmf}\ee
 The  component expansion of the superfields $C,\bar C$ is obtained easily from expansions \eqref{bce},\eqref{bcf} of $E$,$\bar E$ and from 
 \eqref{zh}:   
  \be \bar C=\th  A_3+\bth ( A_1+{\ri}A_2) -\bar\zeta \rho-2{\ri}\th\bth\bar\xi,\label{mbe}\ee
  \be C= \th ( A_1-{\ri}A_2)-\bth A_3-\zeta \rho+2{\ri}\th\bth\xi.\label{mbf}\ee
We finally insert \eqref{cme},\eqref{cmf},\eqref{mbe} and \eqref{mbf} into \eqref{vr} and find the component expansion of the $\uosp$ supersymmetric electrodynamics \eqref{mca}:
$$S(\phi,\Psi,\tilde A,\Xi)=- {\ri} \int\tro  \Psi^\dagger\left( \left(\{M,\Psi\}-2{\ri}  A\Psi+2{\ri}M\rho\Psi\right)- {\ri} \Psi\right)+$$$$+
\int\tro\vert\vert \{M,\phi\} -2{\ri}A\phi+2{\ri}M\rho\phi\vert\vert^2 +2{\ri}\int\tro \left(\bar\phi\Xi^\dagger \Psi-\phi\Psi^\dagger \Xi \right) +$$\be
+ \frac{1}{e^2}\int\tro \left((F(A)+ \rho)^2 +\{y_k,\rho\}\{y_k,\rho\}+ {\ri}\Xi^\dagger\sigma_k\{y_k,\Xi\}+\Xi^\dagger\Xi\right)\label{exp}\ee
 Here 
 \be M=\bma y_3&y_1-{\ri}y_2\\y_1+{\ri}y_2&-y_3 \ema, \quad A\equiv \bma A_3& A_1-{\ri}A_2\\A_1+{\ri}A_2&-A_3\ema,\ee
 and $\Psi,\Psi^\dagger$ are defined by
 \be \Psi:=\bma \psi_-\\\psi_+\ema,\quad \Psi^\dagger =\bma \bar\psi_- &\bar\psi_+\ema.\ee
 Up to a simple renormalisation of the coupling constant, the expression \eqref{exp} contains at the same time the purely bosonic scalar electrodynamics \eqref{act} as well
 as the fermionic electrodynamics (the Schwinger model) in the manifestly $so(3)$ invariant formulation \cite{Jay}.
 We note that the emergence of the  Yukawa-like terms  $\phi\Psi^\dagger \Xi$ is not   specifically inherent to the
 choice of the compact Euclidean "space-time" $S^2$ but it appears also in the flat space version of the supersymmetric
 Schwinger model \cite{Fer}.

In our older paper \cite{Kl1},  we have constructed a different version of the  $\uosp$ supersymmetric electrodynamics on the supersphere than that resumed
by the actions \eqref{mca} or  \eqref{exp}.  The difference in the final component actions is not that big, as we are going to make explicit soon,  nevertheless  from the conceptual point of view the  older construction is very different (and much more complicated)  than the new one.  All difference resides in the gauge field kinetic term: in the new version 
of the theory it  is given by the equation \eqref{gki} and in  the older version \cite{Kl1}  it has also the structure  $\ST\int\std T(\C)^2$, where $T(\C)$ is again the Hermitian super matrix of the $s$-type, but $T(\C)$ is not equal to $\{\cK,\F(\C)\}$ as in \eqref{mca}. Instead, it is given by Eqs. (67) of \cite{Kl1} which can be rewritten in our Poisson language as  
\be T(\C)= \C +{\ri}\{\cM,\C\}+{\ri}\{\C,\cM\}-2\{\cK,\{\cK,\C\}+\{\C,\cK\}\}-2\{\{\cK,\C\}+\{\C,\cK\},\cK\}.\label{hlr}\ee
The reader  may check with the help of Eqs. \eqref{ufo} that $T(\C)$ given by \eqref{hlr} is invariant with respect to the gauge transformation $\C\to\C+\{\cK,\varrho\}$, where $\varrho$ is an arbitrary even real function on $\sus$. Moreover, since the $\uosp$-covariance of $T(\C)$ is also evident, the expression $\ST\int\std T(\C)^2$  has all properties required for an alternative gauge kinetic term. The component expansion based on the ansatz \eqref{zh} then gives
$$\frac{1}{36} \ST\int\std T(\C)^2=$$\be = \int\tro \left(F(A)^2+\frac{2}{9}\rho F(A)+\frac{1}{9}\rho^2 +\{y_k,\rho\}\{y_k,\rho\}+{\ri}  \Xi^\dagger\sigma_k\{y_k,\Xi\}+\frac{1}{9}\Xi^\dagger\Xi
\right). \label{ktn}\ee
By comparing the new gauge kinetic term \eqref{gki} with the old one \eqref{ktn}, we observe that they   coincide up to normalization of certain terms. This circumstance
is extremely favorable because  by taking a suitable  linear combination of the old and the new kinetic terms we can render all gauge fields massless as in the decompactification limit\cite{Fer}. Explicitly, we have
$$\ST\int\std\left(T(\C)^2+4\{\cK,\F(\C)\}^2\right) = $$\be =32 \int\tro \left(F(A)^2 +\{y_k,\rho\}\{y_k,\rho\}+{\ri}  \Xi^\dagger\sigma_k\{y_k,\Xi\} 
\right). \label{fsm}\ee

 \subsection{Fuzzy supersphere} 
   Now we turn to the construction of the supersymmetric electrodynamics on the fuzzy supersphere $\sus_N$. This task was successfully performed 
   in \cite{Kl1} for the "old" gauge kinetic term \eqref{ktn} so here we shall concentrate solely  to the fuzzification of the "new" kinetic term \eqref{gki}.  To begin, recall
    that $\sus_N$ is the noncommutative supermanifold resulting from the quantization of $\sus$ induced by the Poisson brackets \eqref{spb1} or \eqref{pba}. A linear $\uosp$-equivariant quantization map $\Q_N$ associates to smooth superfunctions $f$ on $\sus$ sequences of $(N+1\vert N)\times (N+1\vert N)$ supermatrices $\Q_N(f)$ which are called the fuzzy superfunctions. We shall not need an explicit formula for the quantization map $\Q_N$ but we need three basic properties of it: 
     \be \Q_N(f)\Q_N(g)=\Q_N(fg)+O\biggl(\frac{2}{\sqrt{N^2+N}}\biggr),\label{sp1}\ee
\be [\Q_N(f),\Q_N(g)]= {\ri}\frac{2}{\sqrt{N^2+N}}\Q_N(\{f,g\})+ O\biggl(\frac{4}{N^2+N}\biggr),\label{sp2}\ee
 \be  \frac{1}{2\pi}\int_{\sus}\std f = -\STr (\Q_N(f))+O\biggl(\frac{2}{\sqrt{N^2+N}}\biggr).\label{sp3}\ee
 Obviously the parameter $2/\sqrt{N^2+N}$ plays the role of the Planck constant for the quantization map $\Q_N$.  It is also important to stress that   $\STr$ stands for the supertrace of the $(N+1\vert N)\times (N+1\vert N)$ supermatrices while, throughout this paper, we reserve the symbol $\ST$ 
for the supertrace of $(2\vert 1)\times (2\vert 1)$ supermatrices.

 To give a flavor, what the map $\Q_N$ is about, let us  make explicit  the quantized versions of the   functions  $1$, $w$, $y_k$, $\th$, $\bth$, $\zeta$ and  $\bar\zeta$   on $\sus$:
\be \Q_N(1)=\bma \1_{N+1}&0\\0&\1_N\ema,\quad \Q_N(w)=\bma -\sqrt{\frac{{N}}{{N+1}}}\1_{N+1}&0\\0&-\sqrt{\frac{{N+1}}{{N}}}\1_N\ema,\ee
\be \Q_N(y_k)=\bma \sqrt{\frac{{N+2}}{{N+1}}}Q_{N+1}(x_k)&0\\0&\sqrt{\frac{{N-1}}{{N}}}Q_N(x_k)\ema,\ee
\be \Q_N(\th)=\frac{1}{\sqrt{N^2+N}}\bma 0&T_1\\T^\dagger_2&0\ema,\quad \Q_N(\bth)=\frac{1}{\sqrt{N^2+N}}\bma 0&T_2\\-T^\dagger_1&0\ema,\ee
\be \Q_N(\zeta)=\frac{1}{\sqrt{N^2+N}}\bma 0&-T_2\\-T^\dagger_1&0\ema,\quad \Q_N(\bar\zeta)=\frac{1}{\sqrt{N^2+N}}\bma 0&-T_1\\T^\dagger_2&0\ema.\ee
Here $\1_N$ stands for the unit $N\times N$-matrix, $Q_N(x_k)$ are the quantized generators of the ordinary bosonic fuzzy sphere and the $(N+1)\times N$ matrices $T_1,T_2$ are given by
  \be T_1:=\bma \sqrt{N}&0&\dots &0\\0&\sqrt{N-1}&\dots&0\\0&\dots&\dots&0\\0&\dots&\dots&0\\0&\dots&\sqrt{2}&0\\0&\dots&0&\sqrt{1}\\0&\dots&0&0
    \ema,  T_2:=\bma 0&0&\dots &0\\\sqrt{1}&0&\dots&0\\0&\sqrt{2}&\dots&0\\0&\dots&\dots&0\\0&\dots&\dots&0\\0&\dots&\sqrt{N-1}&0\\0&\dots&0&\sqrt{N}
    \ema.\ee
In particular, it is then easy to verify that it holds the basic  fuzzy supersphere relation:
\be \Q_N(y_1)^2+ \Q_N(y_2)^2+ \Q_N(y_3)^2+\Q_N(\th)\Q_N(\bth)-\Q_N(\bth)\Q_N(\th)=\Q_N(1).\label{sfuz}\ee
 In what follows, we shall adopt a notation keeping the dependence on $N$ tacit:
\be \hat y_k:=\Q_N(y_k), \hat w:=\Q_N(w), \hat \th:=\Q_N(\th), \hat \bth:=\Q_N(\bth),  \hat\zeta:=\Q_N(\zeta),\hat{\bar\zeta}:=\Q_N(\bar\zeta).\label{rep}\ee
It can be straightforwardly checked that  the following   supermatrices $L_k,V,\bar V$
\be L_k:=\frac{\sqrt{N^2+N}}{2}\hat y_k,\quad V:=\frac{\sqrt{N^2+N}}{2}\Q_N(\th), \quad \bar V:=\frac{\sqrt{N^2+N}}{2}\Q_N(\bth),\label{sgn}\ee
realize a $(2N+1)$-dimensional graded unitary representation of the Lie superalgebra $\uosp$. The case
$N=1$ corresponds to the defining representation of $\uosp$ in terms of $(2\vert 1)\times (2\vert 1)$ supermatrices.

In what follows, we shall need the fuzzy versions of the supermatrices $\cM$ and $\cK$. We define them
as 
   \be \hat\cM:=\bma \hat y_3 &\hat y_1-{\ri}\hat y_2 & -\hat\bth\\ \hat y_1+{\ri}\hat y_2& -\hat y_3 & \hat\th\\
    \hat\th & \hat\bth& 0\ema, \quad \hat\cK=\bma \hat w&0&\hat\zeta\\0&\hat w&-\hat{\bar\zeta}\\\hat{\bar\zeta}&\hat\zeta&2\hat w\ema.\label{hmk}\ee
Note that $\hat\cM$ is of the $v$-type while $\hat\cK$ is of $s$-type.  The supermatrices $\hat\cM$ and $\hat\cK$  turn out to fulfil the following identities
which will be useful to show the emergence of the supersymmetric Schwinger model \eqref{mca} as the large $N$ limit
of a certain supermatrix model. Here they are:
\be \hat\cM^2=\frac{N+1/2}{\sqrt{N^2+N}}\hat \cK+2-\frac{3/2}{\sqrt{N^2+N}}\hat\cM,\label{r1}\ee
\be (\hat\cK+2)^2=(\hat\cK+2)+\frac{1}{\sqrt{N^2+N}}\hat\cM+\frac{3}{2}\left(1-\sqrt{\frac{N}{N+1}}\right)\left(1-\sqrt{\frac{N+1}{N}}\right)\hat\cK,\label{r2}\ee
\be \hat\cM\hat\cK=-\frac{N+1/2}{\sqrt{N^2+N}}\hat\cM-\frac{1/2}{\sqrt{N^2+N}}\hat\cK.\label{r3}\ee

\subsection{Supermatrix model}

Now we describe  the construction of the manifestly supersymmetric gauge theory living on the fuzzy supersphere which
in the large $N$ limit yields the supersymmetric electrodynamics \eqref{mca}.
The superfields present in  this noncommutative theory are simply the  $\Q_N$-quantizations of the superfields $\Phi$, $C$, $\bar C$
and $C_0$  living on the ordinary supersphere and we shall denote
them as $\hat\Phi$, $\hat C$, $\hat{\bar C}$ and $\hat C_0$. 
Thus $\hat\Phi$ and $\hat C_0$ will be   even 
$(N+1\vert N)\times (N+1\vert N)$ supermatrices ($\hat C_0$ Hermitian) and $\hat C$,$\hat{\bar C}$ will be odd $(N+1\vert N)\times (N+1\vert N)$ supermatrices Hermitian-conjugated to each other.  As in the commutative case, we arrange
the fuzzy gauge superfields $\hat C_0$, $\hat C$ and $\hat{\bar C}$ into the traceless Hermitian $(2\vert 1)\times (2\vert 1)$ supermatrix $\hat\C$ of the $s$-type:
\be \hat\C:=\bma \hat C_0&0&\hat C\\0&\hat C_0&-\hat{\bar C}\\
\hat{\bar C}&\hat C&2\hat C_0\ema.\ee
We shall require, moreover, that $\hat\C$ obey the following constraint 
\be \ST\left(\hat{ \cK}\hat C+\hat C\hat{ \cK}+\frac{2}{\sqrt{N^2+N}}\hat C\hat C\right)=0.\label{fcon}\ee
Note that the constraint \eqref{fcon} is the
fuzzy analogue of the commutative constraint \eqref{conx} because it follows from \eqref{sp1}:
\be \ST\left(\hat{ \cK}\hat C+\hat C\hat{ \cK}+\frac{2}{\sqrt{N^2+N}}\hat C\hat C\right)=2\Q_N(\ST(\cK\C))+O\left(\frac{2}{\sqrt{N^2+N}}\right).\ee
Here recall that  $\ST$ stands for the supertrace of $(2\vert 1)\times (2\vert 1)$ supermatrices  whereas  the symbol $\STr$ (used e.g. in the next equation)   denotes the supertrace of the $(N+1\vert N)\times (N+1\vert N)$ supermatrices.

Consider now an action 
\be S_N(\hat\Phi,\cP)=-\frac{\pi (N^2+N)}{2}\ST \ \STr  \biggl(( \cP\hat\Phi-\hat\Phi\hat\cK)^\dagger(\cP\hat\Phi-\hat\Phi\hat\cK)+
\frac{1}{e^2}[ \cP, \F(\cP)]^2\biggr),\label{fsa}\ee
where 
\be \cP:=\hat\cK+\frac{2}{\sqrt{N^2+N}}\hat\C, \quad \F(\cP):=-{\ri}\biggl(\frac{N^2+N}{4}\biggr)^{\frac{3}{2}}\ST(\cP^2_v\cP^2_v-\hat\cK^2_v\hat\cK^2_v)\label{two}\ee
and $\cP^2_v$ means the $v$-type part of the supermatrix $\cP^2$ in the sense of the decomposition \eqref{dcp}.

Eqs. \eqref{r1} and \eqref{r2} imply that the expression $\ST(\hat\cK^2_v\hat\cK^2_v)$ commutes with any function on the
fuzzy supersphere $\sus_N$. It hence follows that the action \eqref{fsa} is invariant with respect to a supergauge symmetry 
\be \hat\Phi \to \U\hat\Phi, \quad \cP\to\U\cP\U^\dagger,\ee
where $\U$ is an arbitrary even superunitary   $(N+1\vert N)\times (N+1\vert N)$ supermatrix. In particular, the fuzzy scalar superfield strangth $\F(\cP)$ transforms as
\be \F(\cP)\to \U\F(\cP)\U^\dagger.\ee
In terms of the fuzzy superfield $\hat\C$,  the supergauge transformation takes the following form:
\be \hat \C\to \U\hat\C\U^\dagger-\frac{\sqrt{N^2+N}}{2}[\hat\cK,\U]\U^\dagger.\ee
It can be equally easily checked that the constraint \eqref{fcon}, which can be rewritten as
\be \ST(\cP^2-\hat\cK^2)=0,\ee 
is also supergauge invariant.

Now we study the $\uosp$ supersymmetry of the fuzzy action \eqref{fsa} with respect 
to the  $\uosp$ variations of the superfields $\hat\Phi$ and $\cP$ 
\be \delta_\V\hat\Phi :=-{\ri}[\V_N,\hat\Phi], \quad \delta_\V\cP:= -{\ri}[\V_N\otimes \1_{2\vert 1}+ \1_{N+1\vert N}\otimes \V,\cP].\label{rul}\ee
Here $\1$ stands for the unit supermatrix with the size indicated by the subscript, $\V$ is the element of $\uosp$ viewed
as the $v$-type traceless even Hermitian supermatrix of the size $(2\vert 1)\times(2\vert 1)$ and $\V_N$ is the
Hermitian supermatrix which represents $\V$ in the $(N+1\vert N)$ representation of $\uosp$ described in \eqref{sgn}. 

Restricting a Hermitian supermatrix to  its $v$-part is an operation interchangeable with the $\uosp$ transformation,
hence the supermatrix $\cP^2_v$ transforms as
\be \delta_\V\cP^2_v=-{\ri}[\V_N\otimes \1_{2\vert 1}+ \1_{N+1\vert N}\otimes \V,\cP^2_v]\ee
and $\F(\cP)$ transforms as
\be \delta_\V\F(\cP) =-{\ri}[\V_N,\F(\cP)].\label{peg}\ee
The $\uosp$ supersymmetry of the action \eqref{fsa} now follows easily from \eqref{rul},\eqref{peg}, 
from the cyclic properties of the supertraces $\ST$ and $\STr$ and from the fact that
\be [\V_N\otimes \1_{2\vert 1}+ \1_{N+1\vert N}\otimes \V,\cK]=0.\ee

The last thing to be done is to show that the large $N$ limit of the supermatrix model action \eqref{fsa}
gives the action \eqref{mca} of the supersymmetric electrodynamics on the (graded)commutative supersphere $\sus$. We start by evaluating explicitly the $v$-part of the matrix $\hat\C\hat\cK+\hat\cK\hat\C$ :
\be (\hat\C\hat\cK+\hat\cK\hat\C)_v= \jp\bma [\hat\zeta,\hat{\bar C}]+[\hat{\bar\zeta},\hat{ C}]&2[\hat\zeta,\hat C]&[\hat\zeta,\hat C_0]-[\hat w,\hat C]\\ -2[\hat{\bar\zeta},\hat{\bar C}]&-[\hat\zeta,\hat{\bar C}]-[\hat{\bar\zeta},\hat{ C}]& [\hat w,\hat{\bar C}]-[\hat{\bar\zeta},\hat C_0]\\ [\hat w,\hat{\bar C}] -[\hat{\bar\zeta},\hat C_0]&[\hat w,\hat C]-[\hat\zeta,\hat C_0]&0\ema\label{mtc}\ee
It is important to stress that all commutators appearing in \eqref{mtc} are {\it graded}. Since  the commutator in \eqref{sp2} is also graded, we find
from \eqref{sp2} that
\be (\hat\C\hat\cK+\hat\cK\hat\C)_v= \frac{\ri }{\sqrt{N^2+N}}\Q_N(\{\C,\cK\}+\{\cK,\C\})+O\left(\frac{4}{N^2+N}\right).\label{prm}\ee
From the formula \eqref{r2}, we deduce 
\be (\hat\cK^2)_v=\frac{1}{\sqrt{N^2+N}}\hat\cM.\ee
This fact and the formula \eqref{prm} allow us to find the expansion of $(\cP^2)_v$ in the Planckian constant $2/\sqrt{N^2+N}$:
\be (\cP^2)_v=\frac{1}{\sqrt{N^2+N}}\hat\cM+\frac{2{\ri}}{N^2+N}\Q_N\left(\{\C,\cK\}+\{\cK,\C\}\right)+O\left(\left(\frac{4}{N^2+N}\right)^{\frac{3}{2}}\right). \label{kkk}\ee
By using \eqref{kkk} and \eqref{two}, we immediately infer the expansion of the fuzzy superfield strength $\F(\cP)$ in the Planckian constant:
$$ \F(\cP)=   \frac{\sqrt{N^2+N}}{4\ri }\ST\left((\hat\C\hat\cK +\hat\cK\hat\C)_v\hat\cM+\hat\cM(\hat \C\hat\cK+ \hat\cK\hat\C)_v\right)+ O\left(\frac{2}{\sqrt{N^2+N}}\right)=$$
\be = \jp\Q_N\left(\ST\left(\cM\{\cK,\C\}+\{\C,\cK\}\cM\right)\right) + O\left(\frac{2}{\sqrt{N^2+N}}\right)  .\label{sep} \ee
Then we find from \eqref{ssfs},\eqref{sp2},\eqref{sp3}  and from the first equation of \eqref{two} that the full kinetic term in the fuzzy action \eqref{fsa} expands as
$$ \frac{-\pi(N^2+N)}{2e^2}\ST  \STr   
 [ \cP, \F(\cP)]^2=\frac{2\pi}{e^2} \STr\Q_N(\ST\{\cK,\F(\C)\}^2)+ O\left(\frac{2}{\sqrt{N^2+N}}\right)=$$
 \be = -\frac{1}{e^2}\ST\int\std\{\cK,\F(\C)\}^2+ O\left(\frac{2}{\sqrt{N^2+N}}\right).\label{pd}\ee
In this way we have recovered from the kinetic term of the fuzzy action  in the large $N$ limit the kinetic term of the (graded)commutative action \eqref{mca}.
 
The calculation of the large $N$ limit of the matter kinetic term in \eqref{fsa} is much easier. In fact, the immediate application of  \eqref{sp2}, \eqref{sp3}  and of the first equation of \eqref{two} yields
\be -{\ri}\frac{\sqrt{N^2+N}}{2}(\cP\hat\Phi-\hat\Phi\hat\cK)=\Q_N(\{\cK,\Phi\}-{\ri}\C\Phi)+ O\left(\frac{2}{\sqrt{N^2+N}}\right).\label{dd}\ee
Finally, putting together \eqref{pd}, \eqref{dd}  and exploiting \eqref{sp3}, we conclude that the large $N$ limit
of the fuzzy action \eqref{fsa} is the action \eqref{mca} of the supersymmetric electrodynamics on the (graded)commutative supersphere. Moreover, it can be obtained from \eqref{sp2} and \eqref{sp3}, that  the gauge symmetry and the $\uosp$ supersymmetry of the fuzzy action  induce in the $N\to\infty$ limit
the gauge symmetry and the $\uosp$ supersymmetry of the (graded)commutative action \eqref{mca}.

\section{Discussion of the results and outlook} The reader might have noticed that in the (graded)commutative part
of our work the matter superfield  $\Phi$ was  viewed just as the complex superfunction on the supersphere $\sus$
and not as a section of a nontrivial line bundle over $\sus$. Said in other words, we   did not yet include supervortices
in the formalism. From the physical point of view such inclusion is necessary since the topologically nontrivial
configurations usually play an important role
in the quantum dynamics of electromagnetically interacting matter in two dimensions. Of course, the problem may be circumvented by studying just vortices and not supervortices. This means, in other words,  to expand the manifestly supersymmetric action of the Schwinger models in components and  to promote
the complex scalar boson $\phi$ contained in the superfield $\Phi$ to a section of an appropriate line bundle.
From the mathematical point of
view, however, such a procedure is not very elegant and  the inclusion of supervortices  in a manifestly supersymmetric way  represents actually an intriguing challenge. 

The crucial point to understand is the geometrical status of the multiplet $C_0,C,\bar C$ of the  gauge superfields.  At the first sight
it looks natural to view $C_0,C,\bar C$ as constituent fields of some connection 
{\normalfont\initfamily
\fontsize{4mm}{4mm}\selectfont C}, however, this hypothetical connection must have more constituents then just {\it three} superfields  $C_0,C,\bar C$
because there are in total {\it four}  independent directions on the supersphere (two even and two odd). The problem is that it is not a priori clear how to define covariant derivatives in {\it all }  independent directions
without introducing new dynamical fields into the action. To say the same thing more geometrically, it is not evident how
to complete a partial connection (given by the covariant derivatives in the directions of the Hamiltonian vector
fields $\{\zeta,.\}$, $\{\bar\zeta,.\}$ and $\{w,.\}$ ) into a full connection {\normalfont\initfamily
\fontsize{4mm}{4mm}\selectfont C}.
The usual  trick which works well in the flat space expresses the covariant derivatives in even directions in terms
of the anticommutators of the covariant derivatives in odd directions. However, this methode turns out not to work in the curved 
space. Indeed, we have checked that there is an obstruction to complete the partial connection $C_0,C,\bar C$ to a full connection {\normalfont\initfamily
\fontsize{4mm}{4mm}\selectfont C} in that particular way and, astonishingly enough, that this obstruction can be quantitatively expressed in terms of the
scalar gauge superfield strength $\F(\C)$ given by \eqref{ssfs}! That means, in other words, that only those partial connections $C_0,C,\bar C$ which  have vanishing field strength $\F(\C)$  can be extended to a full connection {\normalfont\initfamily
\fontsize{4mm}{4mm}\selectfont C}!

We believe that, at the present stage, it is wise to postpone the issue of the inclusion of the supervortices into the formalism and to concentrate beforehand onto two other clues capable to shed additional  light on the problem. The
first clue  to follow is noncommutative. As argued by Steinacker in \cite{Ste}, the study of gauge theories on 
the noncommutative spaces can be simpler than on the commutatives ones. In particular, a lot of geometrically involved
concepts like nontrivial fiber bundles, connections, monopole sectors etc. need not be introduced formerly but they arise simply and naturally from the noncommutative formalism \cite{Ste}. We expect that the generalisation of Steinacker's approach to the noncommutative supersymmetric setting may help to contribute  to give a sound  geometrical meaning to the partial connection fields $C_0,C,\bar C$. The second clue consists in closely examining the 
mathematical structure of gauge theories on the sphere with the extended  $N=(2,2)$ supersymmetry and to inspect the geometrical status of their
$N=(1,1)$ contents. 

Needless to say, another problem awaiting a solution consists in calculating a partition function and related dynamical characteristics of the supermatrix model \eqref{fsa} that we have constructed. Whether the fashionable method of localisation can be useful in this context is an open question.


\begin{thebibliography}{99}
 
  \bibitem{BV} A.P. Balachandran and S. Vaidya, Instantons and Chiral Anomaly in Fuzzy Physics,
  Int.J.Mod.Phys. {\bf A16} (2001) 17, hep-th/9910129
 
\bibitem{BC}  F. Benini and S. Cremonesi, Partition functions of $N=(2,2)$ gauge theories on
$S^2$
and
vortices, Commun.Math.Phys. {\bf 334} (2015) 3, 1483,
arXiv:1206.2356 [hep-th]

\bibitem{BPZ}  F. Benini, D. S. Park and P. Zhao,
Cluster algebras from dualities of $2d$ $N=(2,2)$
quiver gauge theories,
arXiv:1406.2699 [hep-th]

\bibitem{CCG} L. Castellani, R. Catenacci and P. A. Grassi, The Geometry of Supermanifolds and New Supersymmetric Actions,
 arXiv:1503.07886 [hep-th]

\bibitem{CC}  C. Closset and S. Cremonesi,
Comments on $N=(2,2)$ Supersymmetry on Two-
Manifolds, JHEP
{\bf 1407}
(2014) 075,
arXiv:1404.2636 [hep-th]


\bibitem{CW} U. Carow-Watamura and S. Watamura, Noncommutative Geometry and Gauge Theory on Fuzzy Sphere,
Commun. Math. Phys. {\bf 212} (2000) 395, hep-th/9801195

\bibitem{DGFL} N. Doroud, J. Gomis, B. Le Floch and S. Lee,
Exact results in $D=2$ supersymmetric
gauge theories, JHEP
{\bf 1305}
(2013) 093,
arXiv:1206.2606 [hep-th]
 
  \bibitem{DG}   N. Doroud and J. Gomis,
Gauge theory dynamics and K\"ahler potential for Calabi-Yau
complex moduli, JHEP {\bf 1312} (2013) 99,
arXiv:1309.2305 [hep-th]

\bibitem{FS}  G. Festuccia and N. Seiberg,  Rigid Supersymmetric Theories in Curved Superspace, JHEP 
{\bf 1106}  (2011) 114 

\bibitem{Fer} S. Ferrara, Supersymmetric Gauge Theories in two Dimensions, Lett. Nuov. Cim. {\bf 13} (1975) 629

 
\bibitem{GGK} E. Gerchkovitz, J. Gomis and Z. Komargodski,
Sphere partition functions and the Zamolodchikov metric,  JHEP {\bf 1411} (2014) 001,
arXiv:1405.7271 [hep-th]


 \bibitem{GF} 
 J. Gomis and B. L. Floch,
$M2$-brane surface operators and gauge theory dualities in
Toda,
arXiv:1407.1852 [hep-th]

\bibitem{GL} J. Gomis and S. Lee,
Exact K\"ahler potential from gauge theory and mirror symmetry,
JHEP
{\bf 1304}
(2013) 019,
arXiv:1210.6022 [hep-th]

\bibitem{GKP}  H. Grosse, C. Klim\v c\'\i k and  P. Pre\v snajder, Field Theory on a Supersymmetric Lattice, Commun. Math. Phys. {\bf 185} (1997) 155 hep-th/9507074
 
\bibitem{HL} K. Hosomichi and S. Lee, Self-dual Strings and 2D SYM,  JHEP {\bf 1501} (2015) 076,
 arXiv:1406.1802 [hep-th] 
 
\bibitem{Ho1} J. Hoppe,  Quantum theory of a massless relativistic surface and a two dimensional bound state problem, PhD. Thesis,
MIT (1982), available at  http://dspace.mit.edu/handle/1721.1/15717

\bibitem{HR} K. Hori and M. Romo,
Exact results in two-dimensional $(2,2)$ supersymmetric gauge
theories with boundary,
arXiv:1308.2438 [hep-th]


\bibitem{IU} S. Iso and H. Umetsu, Note on gauge theory on fuzzy supersphere,  Phys.Rev. {\bf D69} (2004) 105014,   hep-th/0312307;  

\bibitem{IU2} S. Iso and H. Umetsu, Gauge theory on noncommutative supersphere from supermatrix model, Phys.Rev. {\bf D69} (2004) 105003, hep-th/0311005

\bibitem{IKTW} S. Iso, Y. Kimura, K. Tanaka and K. Wakatsuki, Noncommutative Gauge Theory on Fuzzy Sphere from Matrix Model,
Nucl.Phys. {\bf B604} (2001) 121-147, hep-th/0101102
 
\bibitem{Jay} C. Jayewardena, Schwinger model on $S^2$, Helvetica Physica Acta {\bf 61} (1988) 636

\bibitem{Ki} 
  Y.~Kimura,
  Noncommutative gauge theories on fuzzy sphere and fuzzy torus from matrix model,
  Prog.\ Theor.\ Phys.\  {\bf 106}, 445 (2001)
  hep-th/0103192


\bibitem{Kl2} C. Klim\v c\'\i k, Gauge theories on the noncommutative sphere, Commun. Math. Phys. {\bf 199} (1998) 257, hep-th/9710153

\bibitem{Kl1} C. Klim\v c\'\i k,  A nonperturbative regularization of the supersymmetric Schwinger model, Commun. Math. Phys. {\bf 206} (1999) 567, hep-th/9903112 

\bibitem{Kl3} C. Klim\v c\'\i k, On Poisson geometry and supersymmetric sigma models, Mod. Phys. Lett. {\bf A27} (2012) 1250216, arXiv:1204.4654 [hep-th] 

\bibitem{KNP}  D. Karabali, V.P. Nair and  A.P. Polychronakos,  Spectrum of Schrodinger field in a noncommutative magnetic monopole, Nucl.Phys. {\bf B627} (2002) 565-579, hep-th/0111249 
 
\bibitem{Mad} J. Madore, The Fuzzy Sphere, Class. Quant. Grav. {\bf 9} (1992) 69

\bibitem{Pestun}  V. Pestun, Localization of gauge theory on a four-sphere and supersymmetric Wilson 
loops, Commun. Math. Phys. {\bf 313} (2012) 71, arXiv:0712.2824 [hep-th]

\bibitem{RS} V. Rittenberg and V. Scheunert, Elementary construction of graded Lie 
groups, Journal  Math. Phys. {\bf 19}  (1978) 709 
 
 

\bibitem{SaS} I.B. Samsonov and  D. Sorokin, Gauge and matter superfield theories on $S^2$,
   JHEP {\bf 1409}  (2014) 097,  arXiv:1407.6270 [hep-th] 

 
 \bibitem{Sh} E. Sharpe, A few recent developments in $2d$  $(2,2)$ and $(0,2)$ theories,  arXiv:1501.01628 [hep-th]

 \bibitem{Sch} J. Schwinger, Gauge Invariance and Mass II, Phys. Rev. {\bf 128} (1962) 2425

\bibitem{Ste} H. Steinacker, Quantized Gauge Theory on the Fuzzy Sphere as Random Matrix Model, Nucl.Phys. {\bf B679} (2004) 66,  hep-th/0307075
\bibitem{StS}
  H.~Steinacker and R.~J.~Szabo,
  Nonabelian localization for gauge theory on the fuzzy sphere,
  J.\ Phys.\ Conf.\ Ser.\  {\bf 103} (2008) 012017,
  arXiv:0708.4365 [hep-th]
\bibitem{ST} S. Sugishita and S. Terashima,
Exact results in supersymmetric field theories on
manifolds with boundaries, JHEP
{\bf 1311}
(2013) 021,
arXiv:1308.1973 [hep-th]

\bibitem{Yd} B. Ydri, Notes on noncommutative supersymmetric gauge theory on the fuzzy supersphere,
 Int.J.Mod.Phys. {\bf A22} (2007) 5179,  arXiv:0708.3065 [hep-th] 



\end{thebibliography}
\end{document}